\documentclass[a4paper,twoside]{article}

\usepackage{epsfig}
\usepackage{subcaption}

\usepackage{calc}
\usepackage{amstext}
\usepackage{amsthm}
\usepackage{multicol}
\usepackage{pslatex}
\usepackage{apalike}
\usepackage{algorithm2e}
\usepackage[bottom]{footmisc}
\usepackage{graphicx}
\usepackage{todonotes}
\usepackage{amsmath}
\usepackage{amssymb}
\usepackage{caption}
\usepackage{comment}
\usepackage{soul}
\usepackage{hyperref}
\hypersetup{
	colorlinks=true,
	linkcolor={red!50!black},
	citecolor={blue!60!black},
	urlcolor={blue!80!black},
	pdfauthor={Author 1, Author 2, Author 3},
	pdftitle={{Process-Aware Analysis of Treatment Paths in Heart Failure Patients: A Case Study}},
	pdfsubject={Process Mining in Healthcare},
	pdfkeywords={Process Mining, Healthcare, Heart Failure, Patient Cohort Analysis},
	pdfproducer={LaTeX},
	pdfcreator={pdfLaTeX},
	bookmarksopen=true
}

\newtheorem{definition}{Definition} 

\usepackage{rotating}
\usepackage{cleveref}

\newcommand{\hf}[0]{HF}
\newcommand{\hhf}[0]{HHF}
\newcommand{\universe}[0]{\mathcal{U}}

\usepackage{SCITEPRESS}     

\begin{document}
\title{Process-Aware Analysis of Treatment Paths in Heart Failure Patients:\\A Case Study}

\author{
\authorname{Harry H. Beyel\sup{1}\orcidAuthor{0000-0002-6541-3848}, Marlo Verket\sup{2}\orcidAuthor{0000-0002-3361-3722}, Viki Peeva\sup{1}\orcidAuthor{0000-0001-7144-5136}, Christian Rennert\sup{1}\orcidAuthor{0000-0003-4614-6171}, Marco Pegoraro\sup{1}\orcidAuthor{0000-0002-8997-7517}, Katharina Schütt\sup{2}\orcidAuthor{0000-0002-0162-5219}, Wil~M.P.~van der~Aalst\sup{1}\orcidAuthor{0000-0002-0955-6940} and Nikolaus~Marx\sup{2}\orcidAuthor{0000-0001-6141-634X}}
\affiliation{\sup{1}Chair of Process and Data Science (PADS), RWTH Aachen University, Aachen, Germany}
\affiliation{\sup{2}Department of Internal Medicine I, RWTH Aachen University Hospital, Aachen, Germany}
\email{\{beyel, peeva, rennert, pegoraro, vwdaalst\}@pads.rwth-aachen.de, \{mverket, kschuett, nmarx\}@ukaachen.de}
}

\keywords{Process Mining, Healthcare, Heart Failure, Patient Cohort Analysis}

\abstract{Process mining in healthcare presents a range of challenges when working with different types of data within the healthcare domain. There is high diversity considering the variety of data collected from healthcare processes: operational processes given by claims data, a collection of events during surgery, data related to pre-operative and post-operative care, and high-level data collections based on regular ambulant visits with no apparent events. In this case study, a data set from the last category is analyzed. We apply process-mining techniques on sparse patient heart failure data and investigate whether an information gain towards several research questions is achievable. Here, available data are transformed into an event log format, and process discovery and conformance checking are applied. Additionally, patients are split into different cohorts based on comorbidities, such as diabetes and chronic kidney disease, and multiple statistics are compared between the cohorts. Conclusively, we apply decision mining to determine whether a patient will have a cardiovascular outcome and whether a patient will die.}

\onecolumn \maketitle \normalsize \setcounter{footnote}{0} \vfill

\section{\uppercase{Introduction}}

\emph{Heart Failure} (\hf{}) is a complex chronic disease, and the prevalence of HF continues to increase globally. 
\hf{} is one of the main causes of hospitalization for subjects aged 65 years or older, resulting in high costs and a major social impact. 
The management of multimodal treatment for patients with chronic \hf{} is complex.
Despite good treatment, patients with \hf{} still have a poor prognosis, and about 15\% to 25\% die within the first five years after their diagnosis.
In the presence of comorbidities, such as diabetes and chronic kidney disease, this rate rises at least by a 2-fold factor to about $50\%$.~\cite{Romero_Gonz_lez_2020}.
With an abundance of clinical trials focusing on treatment strategies to reduce \emph{Cardiovascular Outcomes} (COs), such as \emph{Hospitalization for Heart Failure} (\hhf{}) and \emph{cardiovascular or all-cause mortality}~\cite{Sattar_2021,Zelniker_2019}, it is still hard to predict which patients respond positively to treatment strategies and which do not. 
A possible solution to gain insights into the treatment responses of patients when data are recorded is \textit{process mining}.\par
The process-mining discipline can generally be split into three main areas: process discovery, conformance checking, and process enhancement~\cite{DBLP:books/sp/Aalst16}. Process-discovery techniques extract a comprehensible process model that represents the underlying behavior in the data. Conformance-checking techniques quantify how well a process model represents the behavior in the recorded data. Process-enhancement techniques can decorate a process model with additional information and insights about the process, e.g., about time, frequency, and explanations of decision points.
While the field of process mining originated in the context of business processes, it has been applied with remarkable success to a number of other disciplines in recent years.
A prominent example of such disciplines is healthcare science~\cite{DBLP:series/sbbpm/MansAV15,DBLP:journals/jbi/Munoz-GamaMFJSH22}.
More specifically, the last few years have seen a surge of process mining analyses of medical data, propelled by advancements in healthcare information systems and by the ever-increasing demand for data-centric solutions to aid in the management of critical situations such as epidemics~\cite{DBLP:conf/bis/0001NBAMM21,DBLP:conf/icpm/Benevento0ABPBA22}.
Applications include the analysis of clinical pathways, patient behavior, and personnel tasks for a large array of diseases~\cite{DBLP:journals/widm/GuzzoRV22}.\par
\begin{figure}[t]
    \centering
    \includegraphics[width=\columnwidth]{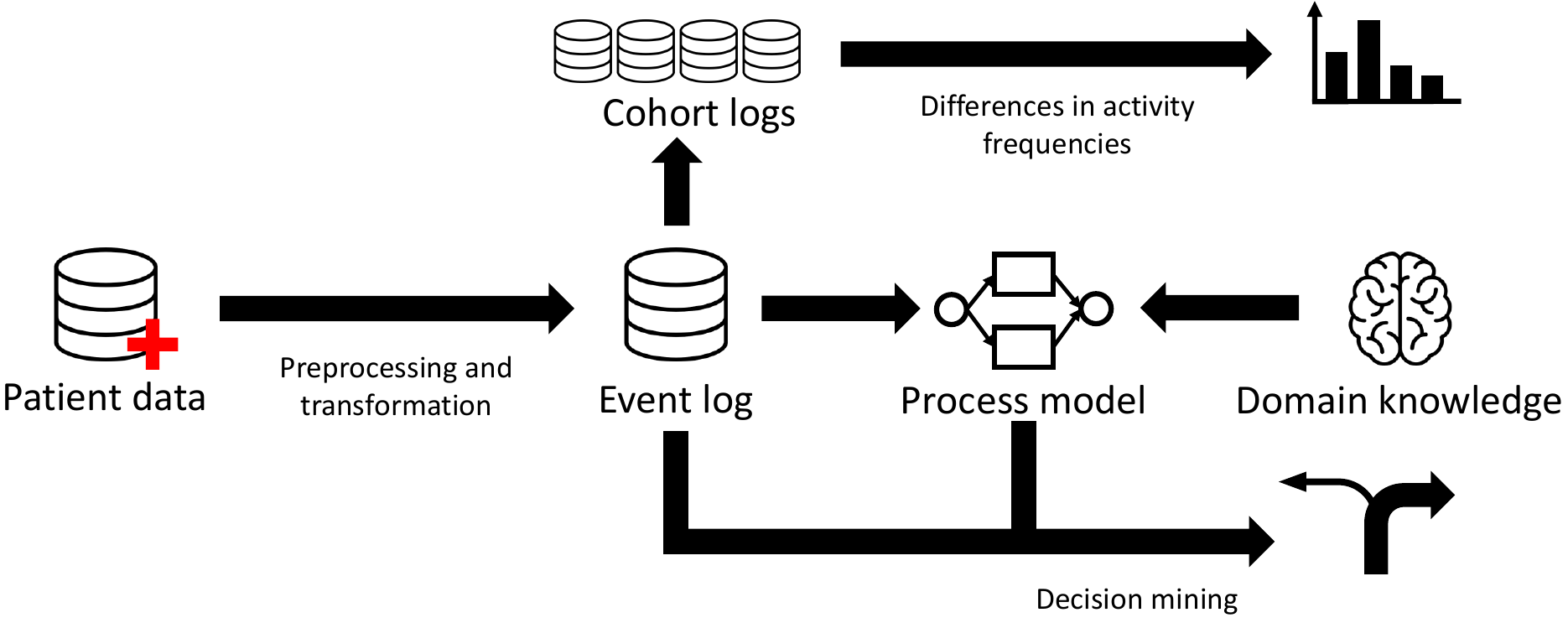}
    \caption{Overview of our work.}
    \label{fig:overview}
\end{figure}
In this paper, we contribute to the literature on process mining in cardiology by reporting our findings on data recorded for patients with HF treated within the Aachen Longitudinal Heart Failure and Diabetes Registry Study (ALIDIA). An overview of our work is shown in~\Cref{fig:overview}.
First, we transform patient data into an event log suitable for process mining. Second, we discover a collection of patients' treatment paths and decide on a representative process model. Third, we compare different subgroups of patients concerning the frequency of events. Fourth, we discover reasons for certain decisions in our process model.\par
The remainder of this paper is organized as follows.
In \Cref{sec:related_work}, we present related work to this study. 
In \Cref{sec:preliminaries}, we lay down preliminary concepts.
Subsequently, in \Cref{sec:datacollection}, we present the collected data.
In \Cref{sec:preprocessing}, we preprocess the available data to select the information we consider in the remainder and transform the resulting data into an event log suitable for process-mining techniques. 
\Cref{sec:paths} analyzes the treatment paths using alignment-based conformance checking.
In \Cref{sec:comparison}, we split the patients into different cohorts, considering whether they suffer from diabetes or chronic kidney disease, and analyze the similarities and dissimilarities between the cohorts.
\Cref{sec:discover} presents an analysis of the decision points in the treatment process.
We report our results together with the challenges we faced analyzing this type of data. 
\Cref{sec:conclusion} concludes the paper, summarizing our findings, pointing out limitations, and providing points for future work.

\section{\uppercase{Related Work}}
\label{sec:related_work}
In~\cite{DBLP:conf/biostec/VriesNGDM17}, a case study is presented showing how to apply process mining to electronic-medical-record data when dealing with sepsis. In~\cite{DBLP:conf/biostec/BackMH20}, patient flows in a surgical ward are mined. The identification of disease trajectory models is investigated in~\cite{DBLP:conf/biostec/KusumaSMJ20}. The authors apply transformations to an event log to later apply process-mining tools. 
Also, process mining has a substantial history in cardiology.
Early attempts~\cite{DBLP:conf/mie/MansSLPCQA08} compared process models for the treatment process of stroke in different hospitals and showed a process model for the pre-hospital process.\par
There has also been a focus on developing models for \hf{}.
Frameworks for the 30-day re-admission risk management and prediction of \hf{} using machine learning, Divide-n-Discover, and applying data mining techniques have been explored and applied. ~\cite{kerexeta2018predicting,roy2014prediction,chin2014divide}. 
Additionally, a regression model for time series using hospital admissions was developed to understand better the impact of environmental factors, such as weather and air quality, from an open dataset on \hf{} ~\cite{artola2019impact}. 
Other factors in a patient's therapy, like the feasibility and usability of devices, have been summarized for telemedicine ~\cite{zweth2018devices}. 
Other systems, like Computer Interpretable Guidelines, have been used to model when a patient, during treatment of a comorbidity, develops \hf{} ~\cite{bottrighi2018general}. Furthermore, the TranSMART tool and the Alfresco platform have been applied to a workflow from patient data collection to data processing in identifying therapeutic targets for \hf{} with preserved ejection fraction ~\cite{almeida2020computational}.
There are several approaches in managing \hf{}, in particular within the timeframe of 30-day re-admission or providing frameworks for biomarkers. This paper aims to fill the gap by attempting to use different factors in the patient pathways, including diabetes and chronic kidney disease as comorbidities. Our goal is to kickstart an underexplored research direction: the analysis of COs in heart failure patients through a combination of process-aware and statistical analysis techniques, with the objective of understanding (and eventually preventing) the insurgence of cardiovascular outcomes.

\section{\uppercase{Preliminaries}}
\label{sec:preliminaries}


In this section, we formally introduce concepts that are the basics of the techniques we introduce later. Given a set $X$, a sequence $\sigma\in X^{*}$ assigns an enumeration to elements of the set. We denote this with $\sigma=\langle\sigma_{1},\dots,\sigma_{n}\rangle$. In the remainder, we refer with $\sigma_{i}$ to the sequence's i-th element.
To concatenate sequences, we use the symbol $\cdot\,$, i.e., $\langle\sigma_{1},\dots, \sigma_{n}\rangle\cdot\langle\sigma'_{1},\dots,\sigma'_{m}\rangle=\langle\sigma_{1},\dots, \sigma_{n},\sigma'_{1},\dots,\sigma'_{m}\rangle$ for $n \in \mathbb{N}$ and $m \in \mathbb{N}$.
Given a set $X$, $\mathcal{B}(X)$ denotes the set of all multisets over set $X$. For example, if $X=\{x, y, z\}$, a possible bag is $[x,x,y]=[x^2,y]$. To combine multisets, we use $\uplus$, for instance, $[x, y]\uplus[x, z]=[x, y, x, z]=[x^2, y, z]$.\par 
To apply process-mining techniques, we need an event log. An event log consists of at least three mandatory attributes: \textit{case identifier}, \textit{activty name}, and \textit{timestamp}. We use $\mathcal{U}_{case}$ as the universe of case identifiers, $\mathcal{U}_{act}$ as the universe of activity names, and $\mathcal{U}_{time}$ as the universe of timestamps. The following defines an event log.
\begin{definition}[Event Log]
	$\mathcal{U}_{ev}$ is the universe of events. $e\in\mathcal{U}_{ev}$ is an event, $\pi_{act}(e)\in\mathcal{U}_{act}$ is the activity of $e$, $\pi_{case}(e)\in\mathcal{U}_{case}$ is the case of $e$, and $\pi_{time}(e)\in\mathcal{U}_{time}$ the timestamp of $e$. An event log $L\subseteq\universe_{ev}$ is a set of events. For simplicity, we assume that other, here non-defined, functions can be applied to the event, resulting in more attributes.
\end{definition}
An example event log is displayed in~\Cref{tab:event_log_example}.
\begin{table}[t]
\caption{Example event log.}
\centering
    \scalebox{0.8}{
\begin{tabular}{|l|l|l|l|l|l|l|}
\hline
\textbf{RowID} & \textbf{Case} & \textbf{Activity}  & \textbf{Timestamp} \\ \hline
1              & 1337            & a                 & 2023-01-21         \\ \hline
2              & 1337            & b                 & 2023-02-15         \\ \hline
3              & 1337            & c                 & 2023-05-05         \\ \hline
4              & 1338            & d                 & 2023-06-01         \\ \hline
5              & 1338            & e                 & 2023-06-19         \\ \hline
6              & 1338            & c                 & 2023-07-20         \\ \hline
\end{tabular}
}
\label{tab:event_log_example}
\end{table}
In the remainder of this work, we use Petri nets as process model representations. Introduction to them is provided in ~\cite{DBLP:books/sp/Aalst16,DBLP:books/sp/Reisig85a}. In this work, we refer to $\tau$-transitions, respectively, silent transitions. In Petri nets, silent transitions provide a flexible and expressive mechanism for modeling various aspects of process behavior. They enable the representation of internal actions, improving Petri nets’ modeling capabilities. Silent transitions can act as synchronization points, waiting for specific conditions to be met before allowing subsequent transitions to fire. In our figures, they are represented as black transitions. Besides, we use conformance-checking techniques to measure a process model's fitness, precision, and generalization scores. ~\cite{DBLP:books/sp/CarmonaDSW18} provide an introduction to these measurements.

\section{\uppercase{Data collection and preparation}}
\label{sec:datacollection}
The data we analyze in this work consists of HF patients treated in the Aachen Longitudinal Heart Failure and Diabetes Registry Study (ALIDIA) between April 2019 and May 2023.
The original data are in a tabular format and contain two disjoint sets of data records for each patient: outpatient clinical visit information and cardiovascular outcome. We joined these two sets, resulting in one concise set capturing the patient data. We define the following Cardiovascular Outcomes (COs): 
\begin{itemize}
    \item \textit{Hospitalization for heart failure}: The day of hospitalization is defined as when the patient is admitted to the emergency room due to clinical manifestations of HF (new or worsening) (HF).
    \item \textit{Hospitalization for myocardial infarction}: The day of hospitalization is defined as when the patient is admitted to the emergency room for non-ST-elevation myocardial infarction or ST-segment elevation myocardial infarction (MI).
    \item \textit{Hospitalization for stroke}: The day of hospitalization is defined as when the patient is admitted to the emergency room for a transient ischemic attack or stroke (Stroke).
    \item \textit{Hospitalization for cardiovascular diseases}: Hospitalisations that were attributed to a cardiovascular disease (CV) cause and not attributed to other previous categories were defined as hospitalization for CV diseases (CV).
    \item \textit{Death due to any cause}: All deaths not attributed to the category of HF were defined as death due to any cause (Death\_AnyCause).
    \item \textit{Death due to heart failure}: When the death occurred in the context of decompensation HF, this death was defined as due to HF (Death\_HF).
\end{itemize}
Moreover, the patient data captures medication information. To evaluate the changes in \hf{} medication, the different substances within a drug class were converted to unity value, using the HF guidelines from the European Society of Cardiology \cite{McDonagh2021}. In the following, we describe the features of the patient data in more detail. $\mathcal{U}_{PatID}$ is the universe of patient ids.
\begin{itemize}
	\item \textit{PatID}: unique identification number allocated to a specific patient ($\mathcal{U}_{PatID}$).
    \item \textit{LVEF}: left ventricular ejection fraction is the percentage of the amount of blood in the left ventricle pumped with each contraction.
    \item \textit{HFrEF}: HF with reduced ejection fraction is defined as LVEF $\leq 40\%$ .
    \item \textit{HFmrEF}: HF with mildly reduced ejection fraction is defined as LVEF between $40\%$ and $49\%$.
    \item \textit{HFpEF}: HF with preserved ejection fraction is defined as LVEF $\geq 50\%$ and symptoms and signs of HF.
    \item \textit{Weight}: biomarker that provides an alert for worsening \hf{}, when rapid weight gain is present.
    \item \textit{HF diagnosis}: the year the patient is diagnosed with HF.
	\item \textit{NT pro-BNP}: biomarker for diagnosing and severity of \hf{}.
	\item \textit{Diabetes}: underlying disease that may augment CO.
    \item \textit{CKD}: chronic kidney disease, an additional underlying disease that may augment CO.
	\item \textit{Outcome}: COs as defined above, covering HF, stroke, MI, CV, death due to any cause, and death due to HF.
    \item \textit{WBC}: white blood cell count, a biomarker for indicating an infection or inflammation.
    \item \textit{hsTNT}: high-sensitivity cardiac troponin T, a biomarker for myocardial injury, including ischemia.
    \item \textit{IL-6}: Interleukin-6 is a biomarker for inflammatory response and is associated with the risk for mortality and cardiovascular outcomes.
    \item \textit{Urea}: a biomarker that indicates the kidney functionality.
    \item \textit{Beta-blocker}: HF medication to reduce mortality and morbidity for patients with HF.
    \item \textit{ACE-I/ARNI}: angiotensin-converting enzyme inhibitors (ACE-I) or an angiotensin receptor-neprilysin inhibitor (ARNI), HF medication to reduce the risks of morbidity and mortality.
    \item \textit{SGLT-2}: sodium-glucose co-transporter 2 inhibitors, HF medication to reduce HF-related hospitalization and CV death.
    \item \textit{MRA}: mineralocorticoid receptor antagonists, HF medication to reduce mortality and the risk of hospitalization for HF.
    \item \textit{Time}: timestamp of record ($\mathcal{U}_{time}$)
\end{itemize}
The collected data consist of 240 different patients and 1000 instance. An example of patient data is depicted in \Cref{tab:pat_data_example}. Each row corresponds to a patient data instance ($\mathcal{U}_{pd}$). An absence of information is denoted with $\bot$.
\begin{table*}[t]
\caption{Example record of patient data.}
\centering
    \scalebox{0.45}{
\begin{tabular}{|l|l|l|l|l|l|l|l|l|l|l|l|l|l|l|l|l|l|l|l|l|l|}
\hline
\textbf{RowID} & \textbf{PatID} & \textbf{LVEF} & \textbf{HFrEF} & \textbf{HFmrEF} & \textbf{HFpEF} & \textbf{Weight} & \textbf{HF diagnosis} & \textbf{NT pro-BNP} & \textbf{Diabetes} & \textbf{CKD} & \textbf{Outcome} & \textbf{WBC} & \textbf{hsTNT}& \textbf{IL-6} & \textbf{Urea} & \textbf{Beta-Blocker} &\textbf{ACE-I/ARNI} & \textbf{SGLT-2} & \textbf{MRA} & \textbf{Timestamp} \\ \hline
1              & 007            & 50            & 0 &0&1&80&2017& 750.5 & 1 &$\bot$ & $\bot$       & $\bot$ & 24.9& 10.5 & 38 &100&50&10&12.5& 2023-02-20         \\ \hline
2              & 007            & 50            & 0 &0&1&80&2017& 750.5 & 1 &$\bot$ & HF
& $\bot$ & 24.9& 10.5 & 38 &100&50&10&12.5& 2023-02-21         \\ \hline
3              & 007            & 50            & 0 &0&1&80&2017& 750.5 & 1 &$\bot$ & Death\_HF     & $\bot$  & 24.9& 10.5 & 38 &100&50&10&12.5& 2023-02-20         \\ \hline
4              & 008            & 10            & 1 &0&0&99& 2012& $\bot$  & $\bot$  &$\bot$ 
&$\bot$       & 10.2   & 24.9& 10.5 & $\bot$  &$\bot$ &$\bot$ &15&$\bot$ & 2023-02-20         \\ \hline
\end{tabular}
}
\label{tab:pat_data_example}
\end{table*}
A formal representation of patient data is presented in the following.
\begin{definition}[Patient Data and Sequences]
$\mathcal{U}_{pd}$ is the universe of patient data instances. $d\in\universe_{pd}$ is a patient data instance. The previously introduced domains and meaning of attributes hold, i.e., $\pi_{PatID}(d)\in\mathcal{U}_{PatID}$ is the patient id of $d$ and $\pi_{time}(d)\in\mathcal{U}_{time}$ the timestamp of $d$. Additionally, $\pi_{LVEF}(d)$ is the LVEF value of $d$, $\pi_{HFrEF}(d)$ signals if the instance $d$ is related to an HFrEF patient, etc. Patient data $P\subseteq\universe_{pd}$ is a set of patient data instances. The data for a patient are stored in $P_{id}=\{p\in P \mid \pi_{PatID}(p)=id\}$. $seq_{id}$ is the sequential representation of $P_{id}$ such that for $seq_{id}=\langle d_1, \dots, d_n\rangle$ of length $n$, the elements $d_1 \in P_{id}, \dots, d_n \in P_{id}$ are sorted by timestamp from earliest to latest, i.e., $\pi_{time}(d_1)\leq \dots\leq \pi_{time}(d_n)$.
\end{definition}

\section{\uppercase{Data Preprocessing and Transformation}}
\label{sec:preprocessing}
To make the data accessible for process mining, we need to preprocess and transform the data.
We pre-process the data by abstracting the level of detail and deriving new attributes based on existing ones.
For example, medication dose changes might be abstracted to medication increase and medication decrease, or the values binned in several categories.
Similarly, the same procedure is done for measured lab values.
After pre-processing the data, we convert our data to an event log to apply process-mining techniques.
This implies identifying case, activity, and timestamp attributes.
Concerning cases, we use the provided patient id attribute.
As a result, each case reflects the treatment path of a patient.
As timestamps, we use the ones provided in the patient data.
Determining the activity attribute is more challenging since it is the only one of the primary attributes (case, activity, timestamp), for which there is a multitude of options to choose from, and choosing one or another would result in different study focuses.
In this work, with the help of domain experts, we decided to focus on a specific subset of COs.
The transformations we present in the rest of this section are made with this goal in mind.
We applied a rule-based approach.
If a datum in the patient data consists of a CO, we use the outcome as an activity.
Otherwise, for each patient, we check whether a visit happens before or after such an outcome.
If it happens before, we assign it as activity ``Visit before CO''; otherwise, we assign ``Visit after CO''. 
We use the previously introduced sequences ($seq_{id}$) for the transformation process. First, we split the sequences into two halves: the part before the first CO and the part containing the first CO.
\begin{definition}[Sequence Before and After Outcome]    
    Let $P\subseteq\mathcal{U}_{pd}$ be a set of patient data instances and $P_{id}$ and $seq_{id}$ be defined as before. $seq_{id}=pre_{id}\cdot post_{id}$, with $seq_{id}=\langle d_1, \dots, d_i\rangle\cdot\langle d_{i+1}, \dots, d_n\rangle$, such that for all $d_j\in pre_{id}$, $j\in\{1, \dots, i\}$, $\pi_{Outcome}(d_j)=\bot$ and $\pi_{Outcome}(d_{i+1})\neq\bot$.
\end{definition}
For our example data (\Cref{tab:pat_data_example}), there are the following sequences: $pre_{007}=\langle 1\rangle$, $post_{007}=\langle 2, 3\rangle$, $pre_{008}=\langle 4\rangle$, $post_{008}=\langle~\rangle$.\par
In the following, we transform each datum in a sequence before a CO into an event.
\begin{definition}[Transformation Before Outcome]
    Let $P\subseteq\mathcal{U}_{pd}$ be a set of patient data instances and $P_{id}$ and $seq_{id}$ and $pre_{id}$ be defined as before. Let $L\subseteq\mathcal{U}_{ev}$ be an event log. There exists a function $trans_{pre}$ that maps each element of the sequence $pre_{id}$ of length $n$, $d_i \in pre_{id}$, $i\in\{1, \dots, n\}$, to an event such that the following holds:
    \begin{itemize}
        \item $\pi_{PatID}(d_i)=\pi_{case}(trans_{pre}(d_i))$
        \item $\text{``Visit before CO''}=\pi_{act}(trans_{pre}(d_i))$
        \item $\pi_{time}(d_i)=\pi_{time}(trans_{pre}(d_i))$
    \end{itemize}
    Moreover, the values for the different patient data attributes for all $d_i\in pre_{id}$ are the same for the mapped event.
\end{definition}
Finally, we transform each datum in a sequence after a CO into an event.
\begin{definition}[Transformation After Outcome]
     Let $P\subseteq\mathcal{U}_{pd}$ be a set of patient data instances and $P_{id}$ and $seq_{id}$ and $post_{id}$ be defined as before. Let $L\subseteq\mathcal{U}_{ev}$ be an event log. There exists a function $trans_{post}$ that maps each element of the sequence $post_{id}$ of length $n$, $d_i \in post_{id}$, $i\in\{1, \dots, n\}$, to an event such that the following holds:
    \begin{itemize}
        \item $\pi_{PatID}(d_i)=\pi_{case}(trans_{post}(d_i))$
        \item $\text{``Visit after CO''}=\pi_{act}(trans_{post}(d_i))$ if $\pi_{Outcome}(d_i)=\bot$
        \item $\pi_{Outcome}(d_i)=\pi_{act}(trans_{post}(d_i))$ if $\pi_{Outcome}(d_i)\neq\bot$
        \item $\pi_{time}(d_i)=\pi_{time}(trans_{post}(d_i))$
    \end{itemize}
    Moreover, the values for the different patient data attributes for all $d_i\in post_{id}$ are the same for the mapped event.
\end{definition}
We apply these steps to all patients contained in the data.
The result of applying these steps to the patient data portrayed in~\Cref{tab:pat_data_example} is displayed in~\Cref{tab:event_log_example_transfomed}.
\begin{table*}[t]
\caption{Example transformed event log.}
\centering
    \scalebox{0.8}{
\begin{tabular}{|l|l|l|l|l|l|l|l|l|l|l|l|l|l|l|l|l|l|l|l|l|l|l|}
\hline
\textbf{RowID} & \textbf{Case} & \textbf{Activity}  &  \textbf{LVEF}& \textbf{HFrEF} & \textbf{HFmrEF} & \textbf{HFpEF} & $\dots$   & \textbf{MRA} & \textbf{Timestamp} \\ \hline
1              & 007            & Visit before CO   & 50            & 0 &0&1                                            & $\dots$   & 12.5& 2023-02-20         \\ \hline
2              & 007            & HF                & 50            & 0 &0&1                                            & $\dots$   &12.5  & 2023-03-14         \\ \hline
3              & 007            & Death\_HF         & 50            & 0 &0&1                                            & $\dots$   &12.5 & 2023-04-15         \\ \hline
4              & 008            & Visit before CO   & 10            & 1 &0&0                                            & $\dots$   &$\bot$    & 2023-06-18         \\ \hline
\end{tabular}
}
\label{tab:event_log_example_transfomed}
\end{table*}
Applying these steps to our patient data leads to an event log with 240 cases (i.e., patients) and 1000 events. 

\section{\uppercase{Obtaining Treatment Paths}}
\label{sec:paths}
In this section, we investigate the nature and behavior of the treatment paths of patients obtained in the previous section.
To observe the treatment paths, we use process models, in particular, the aforementioned Petri nets.
To validate whether the received process models are representative and the data is compliant, we apply conformance-checking techniques. 
In particular, we measure fitness by utilizing alignments with a standard cost function \cite{DBLP:conf/acsd/AdriansyahSD11}, computed precision \cite{DBLP:journals/isem/AdriansyahMCDA15} and generalization \cite{DBLP:journals/ijcis/BuijsDA14} scores, as well as the simplicity of the model \cite{DBLP:conf/bpm/WeerdtBVB10}.
In addition, we computed F1 scores using the obtained fitness and precision scores.
To receive process models, we applied several process discovery algorithms to the transformed event log, all implemented in ProM\footnote{Available at \url{https://promtools.org/}.}.
Moreover, we consulted domain experts to construct a de-jure model.
The conformance-checking results are shown in \Cref{tab:conformance}. 
\begin{table*}[t]
    \centering
    \small
    \caption{Conformance checking results.}
    \label{tab:conformance}
    \scalebox{0.8}{
    \begin{tabular}{|l|l|l|l|l|l|l|}
    \hline
        \textbf{Algorithm} & \textbf{Parameters} & \textbf{Fitness} & \textbf{Precision} & \textbf{Generalization} & \textbf{Simplicity} & \textbf{F1 score} \\ \hline
        Alpha & - & 0.39 & 0.44 & 0.70 & 0.85 & 0.41 \\ \hline
        Alpha+ & - & 0.39 & 0.44 & 0.70 & 0.85 & 0.41 \\ \hline
        Alpha++ & - & - & - & - & - & - \\ \hline
        Alpha\# & - & 0.96 & 0.57 & 0.78 & \textbf{1.00} & 0.72 \\ \hline
        Directly-follows Miner & 0.0 (paths)& - & - & - & - & - \\ \hline
        Directly-follows Miner & 0.1 (paths) & 0.75 & \textbf{1.00}&\textbf{0.89} & 0.78 & 0.86 \\ \hline
        Directly-follows Miner & 0.2 (paths)& 0.75 & \textbf{1.00} &\textbf{0.89}& 0.78 & 0.86 \\ \hline
        Directly-follows Miner & 0.3 (paths)& 0.75 & \textbf{1.00} & \textbf{0.89} & 0.78 & 0.86 \\ \hline
        Directly-follows Miner & 0.4 (paths)& 0.75 & \textbf{1.00} & \textbf{0.89} & 0.78 & 0.86 \\ \hline
        Directly-follows Miner & 0.5 (paths)& 0.75 & \textbf{1.00} & \textbf{0.89} & 0.78 & 0.86 \\ \hline
        Directly-follows Miner & 0.6 (paths)& 0.75 & \textbf{1.00} & \textbf{0.89} & 0.78 & 0.86 \\ \hline
        Directly-follows Miner & 0.7 (paths)& 0.84 & 0.98 & 0.86 & 0.68 & 0.90 \\ \hline
        Directly-follows Miner & 0.8 (paths)& 0.91 & 0.96 & 0.84 & 0.63 & 0.93 \\ \hline
        Directly-follows Miner & 0.9 (paths)& 0.97 & 0.94 & 0.76 & 0.52 & \textbf{0.95} \\ \hline
        Directly-follows Miner & 1.0 (paths)& \textbf{1.00} & 0.72 & 0.48 & 0.47 & 0.83 \\ \hline
        Inductive Miner complete & - & \textbf{1.00} & 0.62 & 0.80 & 0.64 & 0.77 \\ \hline
        Inductive Miner frequent & 0.0 & \textbf{1.00} & 0.63 & 0.80 & 0.63 & 0.77 \\ \hline
        Inductive Miner frequent & 0.1 & 0.99 & 0.59 & 0.82 & 0.61 & 0.74 \\ \hline
        Inductive Miner frequent & 0.2 & 0.96 & 0.55 & 0.82 & 0.60 & 0.70 \\ \hline
        Inductive Miner frequent & 0.3 & 0.46 & 0.74 & 0.77 & 0.62 & 0.57 \\ \hline
        Inductive Miner frequent & 0.4 & 0.45 & 0.74 & 0.77 & 0.64 & 0.56 \\ \hline
        Inductive Miner frequent & 0.5 & 0.41 & 0.82 & 0.77 & 0.70 & 0.55 \\ \hline
        Inductive Miner frequent & 0.6 & 0.41 & 0.82 & 0.77 & 0.70 & 0.55 \\ \hline
        Inductive Miner frequent & 0.7 & 0.33 & 0.79 & 0.75 & 0.78 & 0.47 \\ \hline
        Inductive Miner frequent & 0.8 & 0.33 & 0.79 & 0.75 & 0.78 & 0.47 \\ \hline
        Inductive Miner frequent & 0.9 & 0.33 & 0.79 & 0.75 & 0.78 & 0.47 \\ \hline
        Inductive Miner frequent & 1.0 & 0.27 & 0.67 & 0.66 & 0.90 & 0.38 \\ \hline
        De-jure model & - & \textbf{1.00} & 0.56 & 0.79 & 0.54 & 0.72 \\ \hline
    \end{tabular}
    }
\end{table*}

\begin{figure}[t]
    \centering
    \includegraphics[width=\columnwidth]{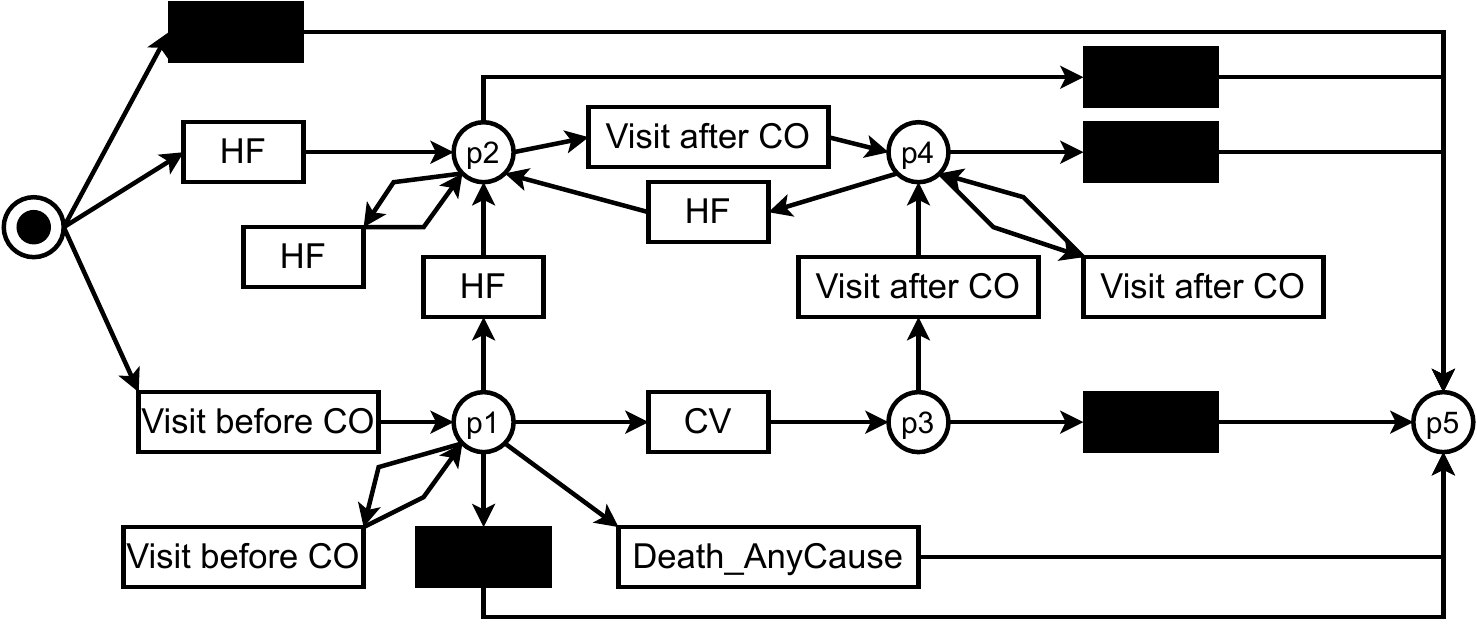}
    \caption{De-facto model of treatment paths.}
    \label{fig:de-facto-model_df}
\end{figure}

\begin{figure}[t]
    \centering
    \includegraphics[width=\columnwidth]{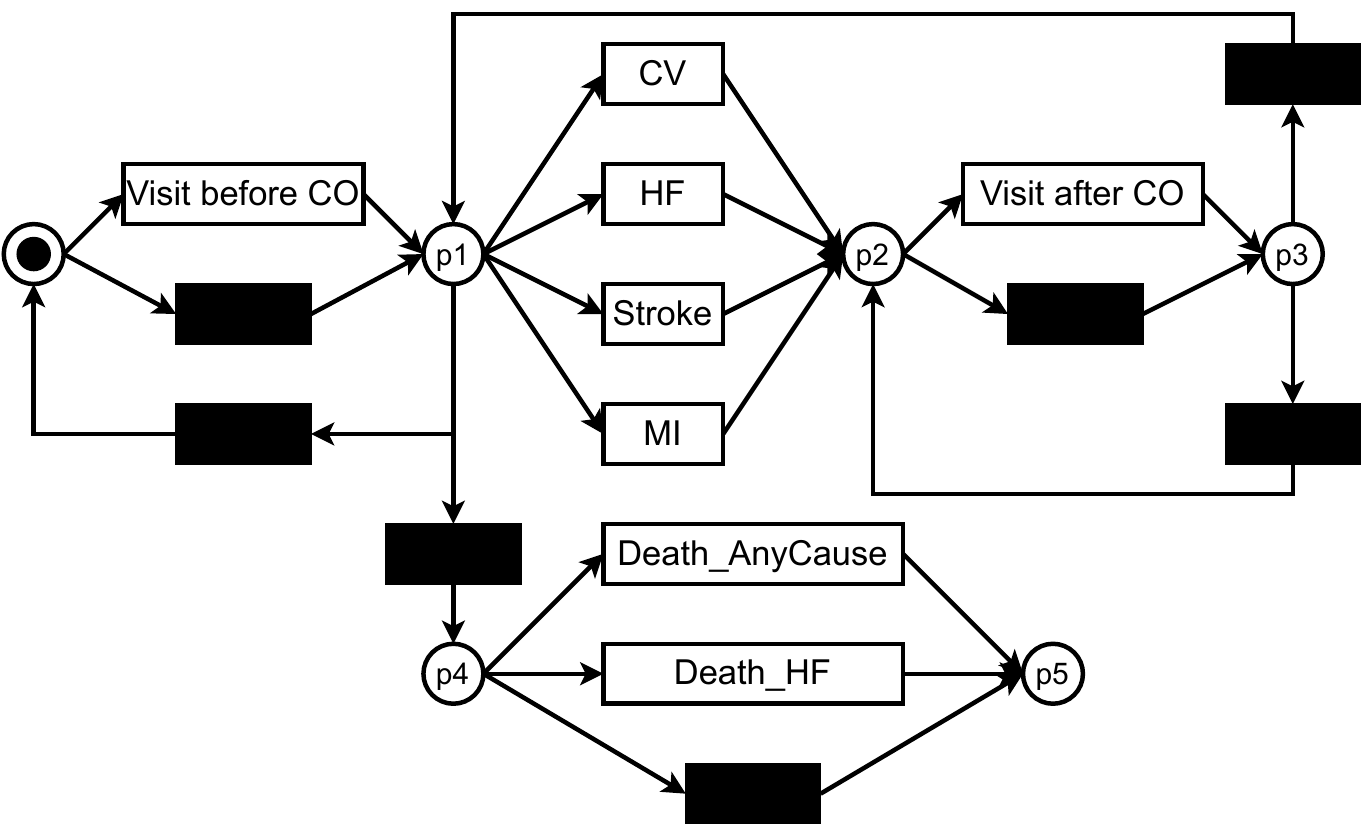}
    \caption{De-jure model of treatment paths.}
    \label{fig:de-facto-model}
\end{figure}

As denoted, the process-discovery algorithms yield various results.
Four models have a fitness score of $1.0$, and six models have a precision score of $1.0$.
The best generalization score is $0.89$, achieved by six models.
One model has a simplicity score of $1.0$, far better than the second-best score ($0.9$).
Concerning the F1 score, the best score is $0.95$, achieved by the model discovered by the directly-follows miner using all activities and $90\%$ of paths. This model is depicted in \Cref{fig:de-facto-model_df}.
However, this model has some issues. First, not all activities are represented as activities. ``Death\_HF'', ``MI'', and ``Stroke'' are missing. Second, the model does not show the possible paths as given in the data. For example, a patient that has an HF (\textit{p2}) can never die. Third, the structure of the process model limits us in applying process-enhancement techniques such that conclusions for domain experts are possible.
As a result, we use the de-jure model.
The advantage of this model is its structure, allowing for decision-mining applications interesting for domain experts.
The model is depicted in \Cref{fig:de-facto-model}.
In the following, we describe the behavior captured in the model.
Starting from the initial marking, there is a choice between having a visit or not.
Independently from the choice, the next place is $p1$.
In $p1$, there are multiple choices.
First, one of the cardiovascular outcomes may happen. 
Second, a visit may happen again, potentially more than two times.
Third, the record of patients ends, either with their death or there is no more record ($p4$).
Firing one of the cardiovascular outcomes leads to place $p2$.
Next, there is a choice between a visit or no visits, leading to place $p3$.
In place $p3$, there is a choice between having another visit or switching to place $p1$.

\section{\uppercase{Cohort Comparison}}
\label{sec:comparison}
In this section, we compare different patient cohorts of our data. First, we present how we split up the data and provide reasons. Second, we discover statistically significant differences concerning the amount of activity frequencies. 
\subsection{Dividing the Data}
We divided the patient population based on three characteristics:
(1) which \hf{} phenotype the patient initially exhibited;
(2) whether the patient had diabetes or not;
(3) whether the patient had chronic kidney disease (CKD) or not.
Patients with \hf{} can be divided into three groups: patients with reduced (HFrEF), mildly reduced (HFmrEF), or preserved (HFpEF) left ventricular ejection fraction (LVEF). 
All phenotypes are based on LVEF with the presence of signs and symptoms of \hf{}.
As introduced earlier, HFrEF is classified with LVEF $\leq$$40\%$, HFmrEF is defined with LVEF between $40\%$ and $49\%$, and HFpEF is with LVEF~$\geq$$50\%$~\cite{McDonagh2021}.
This grouping allows for tailored treatment strategies. 
CKD is defined as showing at least one marker of kidney damage or persistently reduced estimated glomerular filtration (eGFR) rate of $<$$60$ ml/min per $1.73 m^2$ for $>$$3$ months~\cite{levin2013kidney}.
CKD and \hf{} are concurrent diseases that accelerate the progression of outcomes, thus leading to an increase in risk for hospitalization and death.
Furthermore, diabetes is defined as HbA1c value $\geq 6.5\%$ ($\geq 48$ mmol/mol), or when the patient is treated with anti-diabetic medication.
There are two types of diabetes: Type 1 and Type 2.
Over $90\%$ of cases are Type 2 diabetes.
Similarly to CKD, patients with \hf{} and diabetes have a higher risk of cardiovascular mortality, including death due to worsening \hf{}, compared to those without diabetes.
We exclude patients with Type 1 diabetes.

\subsection{Differences in Frequencies of Activities}
First, we count the occurrences for each activity in a given case (i.e., a patient treatment path).
\begin{definition}[Occurrence count: case]
    Let $L\subseteq\mathcal{U}_{ev}$ be an event log, $act\in\mathcal{U}_{act}$ be an activity, and $case\in\mathcal{U}_{case}$ be a case. $count_{c}$ is a function that counts how often an activity occurs in a given case for a given log, i.e., $count_{c}(act, case, L)=|\{e\in L\mid\pi_{case}(e)=case \land \pi_{act}(e)=act\}|.$
\end{definition}
In the next step, we perform this action on all cases of a given event log, resulting in a multiset, in which the cardinality of each element denotes the number of occurrences of the given activity in a case.
\begin{definition}[Occurrence count: event log]
    Let $L\subseteq\mathcal{U}_{ev}$ be an event log and $act\in\mathcal{U}_{act}$ be an activity. Then $C_{L}=\{\pi_{case}(e)\mid e\in L\}$ provides all cases of an event log $L$. $count_{l}$ is a function that returns a multiset of case-wise occurrences of an activity in an event log, i.e., $count_{l}(act, L)=\biguplus_{case\in C_{L}}[count_{c}(act, case, L)]$.
\end{definition}
Applying this on the event log shown in~\Cref{tab:event_log_example_transfomed} concerning activity ``Visit before CO'' leads to $[1 ,1]$, concerning activity ``HF'' leads to $[1, 0]$.\par
We apply this methodology considering all activities. We consider two greater cohorts: patients who have or do not have diabetes and different LVEF categories (HFrEF, HFmrEF, HFpEF); and patients who have or do not have CKD and different LVEF categories (HFrEF, HFmrEF, HFpEF). Each cohort consists of six groups. We applied the Kruskal-Wallis test \cite{kruskal} to check for differences between groups for each cohort. The Kruskal-Wallis test is a non-parametric version of the ANOVA test \cite{Fisher1992}. If there is a statistically significant difference in the number of occurrences, we applied Dunn's test \cite{dunn} with Bonferroni adjustment to check which group differs.\par
The results concerning Kruskal-Wallis tests \cite{kruskal} are shown in~\Cref{tab:kruskal_wallis_diabetes,tab:kruskal_wallis_ckd}. 

\begin{table}[t]
    \centering
    \caption{P-values after applying Kruskal-Wallis test \cite{kruskal} concerning diabetes (yes or no) and different LVEF categories (HFrEF, HFmrEF, HFpEF), resulting in six groups for each activity.}
    \begin{tabular}{|l|c|}\hline
        \textbf{Activity}   & \textbf{p-value}\\ \hline
        Visit before CO& 0.151 \\\hline
        Visit after CO&  0.566\\\hline
        Hosp\_CV&0.485\\\hline
        Hosp\_HF&0.205\\\hline
        Hosp\_Stroke&0.841\\\hline
        Hosp\_MI&0.225\\\hline
        Death\_AnyCause&0.137\\\hline
        Death\_HF&\textit{0.017}\\\hline
    \end{tabular}
    \label{tab:kruskal_wallis_diabetes}
\end{table}

\begin{table}[t]
    \centering
    \caption{P-values after applying Kruskal-Wallis test \cite{kruskal} concerning CKD (yes or no) and different LVEF categories (HFrEF, HFmrEF, HFpEF), resulting in six groups for each activity.}
    \begin{tabular}{|l|c|}\hline
        \textbf{Activity}   & \textbf{p-value}\\ \hline
        Visit before CO& 0.550 \\\hline
        Visit after CO&  0.549\\\hline
        Hosp\_CV&0.376\\\hline
        Hosp\_HF&0.113\\\hline
        Hosp\_Stroke&0.949\\\hline
        Hosp\_MI&0.213\\\hline
        Death\_AnyCause&0.287\\\hline
        Death\_HF&0.183\\\hline
    \end{tabular}
    \label{tab:kruskal_wallis_ckd}
\end{table}
We consider a difference statistically significant if a p-value is smaller than 0.05. In~\Cref{tab:kruskal_wallis_diabetes,tab:kruskal_wallis_ckd}, such a value can only be noted in ~\Cref{tab:kruskal_wallis_diabetes}, when considering the activity `Death\_AnyCause'', diabetes (yes or no), and different LVEF categories (0.017). As a result, Dunn's test \cite{dunn} with Bonferroni adjustment is applied to observe which groups differ. The result is displayed in~\Cref{tab:dunn_result}.
\begin{table*}[t]
    \centering
    \small
    \caption{Adjusted p-values after applying Dunn's test \cite{dunn} with Bonferroni adjustment concerning the activity ``Death\_AnyCause'', diabetes (D, yes (1) or no (0)), and different LVEF categories.}
    \scalebox{0.8}{
    \begin{tabular}{|l|l|l|l|l|l|l|}\hline
            & \textbf{D=0 and HFmrEF} & \textbf{D=0 and HFpEF}  & \textbf{D=0 and HFrEF} & \textbf{D=1 and HFmrEF} & \textbf{D=1 and HFpEF}  & \textbf{D=1 and HFrEF}    \\\hline
         \textbf{D=0 and HFmrEF}  & 1.00  & 1.00   &1.00  & 1.00   & 1.00  & 0.45                              \\\hline
         \textbf{D=0 and HFpEF}   & 1.00  & 1.00   &1.00  & 1.00   & 1.00  & 0.11                           \\\hline
         \textbf{D=0 and HFrEF}   & 1.00  & 1.00   &1.00  & 1.00   & 1.00  &\textit{0.01}                             \\\hline
         \textbf{D=1 and HFmrEF}  & 1.00  & 1.00   &1.00  & 1.00   & 1.00  & 0.36                             \\\hline
         \textbf{D=1 and HFpEF}   & 1.00  & 1.00   &1.00  & 1.00   & 1.00  & 1.00                             \\\hline
         \textbf{D=1 and HFrEF}   & 0.45  & 0.11   &\textit{0.01}  & 0.36   & 1.00  & 1.00                            \\\hline
    \end{tabular}
    }
    \label{tab:dunn_result}
\end{table*}
As it can be noted, the group having diabetes and being associated with HFrEF is different from nearly all other groups. However, a stronger statistically significant difference can be noted between the mentioned group of patients and patients having no diabetes and being part of the HFrEF group. As a result, ``Death\_AnyCause'' happens statistically significantly more often for HFrEF patients if they have diabetes. 

\section{\uppercase{Investigating Reasons}}
\label{sec:discover}
In this section, we investigate the reasons for decisions in the model.
To investigate the reasons, two things are required: the transformed event log and at least one decision point. A decision point is a place in a Petri net with more than one outgoing arc.
There are two interesting decision points in the model shown in~\Cref{fig:de-facto-model}: places $p1$ and $p4$.
The former is a decision point concerning COs.
The latter points out the decision between two death options and the end of a patient's record.
Using the event log, we use the information associated with the events to discover the reasons behind the decision.
Since HFrEF patients are in a worse state than the others, we additionally applied our approach to records containing patients with that characteristic separately.
The probability distribution for the outcomes of the mentioned places considering the whole population and the HFrEF patients is displayed in~\Cref{tab:distribution_deicison_mining}. We summarised the execution of $\tau$-transitions as ``None''.
\begin{table*}[t]
    \centering
    \small
    \caption{Distribution of next activities for places \textit{p1} and \textit{p4}.}
    \scalebox{0.8}{
    \begin{tabular}{|l||l|l|l|l|l|l|l|l|l|}\hline
        \textbf{Place and population}   & \textbf{None}   & \textbf{HF}  & \textbf{CV}   &\textbf{Stroke}  &\textbf{MI}  &\textbf{Death\_AnyCause} &\textbf{Death\_HF} \\\hline
        Place \textit{p1}(all patients) &91.86&5.78&1.90&0.30&0.15 &-&-\\\hline
        Place \textit{p1} (HFrEF patients)&91.30 &6.10&2.10 &0.40&0.10&-&-\\\hline
        Place \textit{p4} (all patients)& 98.29&-& -& - & - & 1.39 & 0.33\\\hline
        Place \textit{p4} (HFrEF patients)& 97.89&-& -& - & - & 1.71 & 0.40\\\hline
    \end{tabular}
    }
    \label{tab:distribution_deicison_mining}
\end{table*}
The results of applying various classification techniques are displayed in~\Cref{tab:decision_mining}.
\begin{table*}[t]
    \centering
    \small
    \caption{Highest measured accuracy of prediction measured using multiple strategies for different places and patient populations in the model shown in \Cref{fig:de-facto-model}.}
    \scalebox{0.8}{
    \begin{tabular}{|l||l|l|l|l|l|l|l|l|l|}\hline
        Place and population   & \begin{tabular}[c]{@{}l@{}}Naive\\Bayes\end{tabular}   & \begin{tabular}[c]{@{}l@{}}Generalized\\ Linear\\ Model\end{tabular}  & \begin{tabular}[c]{@{}l@{}}Logistic\\ Regression\end{tabular}   &\begin{tabular}[c]{@{}l@{}}Fast\\ Large\\ Margin\end{tabular}  &\begin{tabular}[c]{@{}l@{}}Deep\\ Learning\end{tabular}  &\begin{tabular}[c]{@{}l@{}}Decision\\Tree\end{tabular}  &\begin{tabular}[c]{@{}l@{}}Random\\ Forest\end{tabular}  &\begin{tabular}[c]{@{}l@{}}Gradient\\ Boosted\\ Trees\end{tabular} &\begin{tabular}[c]{@{}l@{}}Support\\ Vector\\ Machine\end{tabular}\\ \hline
        Place \textit{p1} (all patients)  & 91.8  & 91.8  & - &-     &91.8    &91.8   &91.8   & 91.8  &91.8 \\\hline
        Place \textit{p1} (HFrEF patients) & 91.4  &86.3   &-  &-     &91.0    &91.4   &88.4   & 91.4  &91.4\\\hline
        Place \textit{p4} (all patients)& 98.0&98.0& -& 98.0&98.0&98.0&98.0&98.0&98.0\\\hline
        Place \textit{p4} (HFrEF patients) &98.6&98.6&-&98.6&98.6&98.6&98.6&98.6&98.6\\\hline
    \end{tabular}
    }
    \label{tab:decision_mining}
\end{table*}
In the following, we investigate the different decision points in greater detail.
\subsection{Reasons for Cardiovascular Outcomes}
COs are highly interesting events in the strategy for patients' treatment.
Understanding why they happen can lead to improvements in the treatment process, such that these outcomes are avoided.
The choice between COs is taking place in place $p1$ in the Petri net depicted in~\Cref{fig:de-facto-model}.
As shown in~\Cref{tab:distribution_deicison_mining}, a CO activity rarely happens. Considering the whole population, a non-CO activity is executed 91.86\% of the time. The algorithms adopt this behavior, conduct a majority vote, and achieve similar accuracy, as shown in~\Cref{tab:decision_mining}. When we consider only HFrEF patients, we observe that 91.3\% of the time, a non-CO activity is executed. At the same time, the accuracy of the algorithms is slightly better, with an accuracy of 91.4\% (naive Bayes, decision trees, gradient-boosted trees, and support vector machines). However, the difference is negligible. We conclude that predicting a CO activity, given our data, is not possible. The distribution over the different classes is skewed, and sampling was infeasible based on the limited data.
\subsection{Reasons for Death}
As stated, in place $p4$ in the model depicted in~\Cref{fig:de-facto-model}, the decision between a patient's death and the end of a patient's records takes place.
In the following, we analyze the reasons for the whole population and HFrEF patients. Considering the distribution of activities, as shown in~\Cref{tab:distribution_deicison_mining}, a death activity happens rarely. Considering the whole population, 98.29\% of the time, the records of patients end without death. However, as shown in~\Cref{tab:decision_mining}, the accuracy is at most 98\%. When considering only HFrEF patients, 97.89\% of the time, the records of patients end without death. The classifiers have an accuracy of 98.6\%, which is better than the majority vote and our best offset to the greatest distribution value. However, given the unbalanced classes, the accuracy is still not satisfying. As for determining reasons for COs, we do not have enough information on all classes.

\section{\uppercase{Conclusion}}
\label{sec:conclusion}

Our work aimed to analyze the reasons for COs and death during the treatment pathway of a patient with HF. In general, we applied process mining-based methodology to patient data for HF in a real-world setting. We transformed patient data into an event log. Subsequently, we applied process-mining techniques to the transformed event log. We discovered process models and applied conformance-checking techniques to the discovered model and a de-jure model. We investigated differences between different patient cohorts. Finally, we performed decision mining using the transformed event log and the de-jure model.\par
However, the results have to be treated with caution. First, the number of patients included is limited.
The records that we analyzed were for 240 patients.
Thus, we analyzed 1000 events, i.e., roughly four events per patient in roughly four years.
Second, patients' data are usually sparse, and the dataset we analyzed is no exception. Types of sparseness include the temporal sparseness that was previously mentioned, but there is also an absence of information for some attributes.
Third, patients' data and their treatment are highly complex. The data we analyzed only cover information recorded at the RWTH Aachen University Hospital. Information about the treatments in other hospitals is limited. Thus, only CV-related hospitalizations from other hospitals are included.\par
This case study allows for open questions and directions for future research in the context of healthcare. First, the used activities are generic. The data consists of biomarkers, which can be used to generate more meaningful activity names. Second, for some visits, there exist textual reports. Using text-mining techniques, information can be extracted from these reports, enabling a more precise description of the record. Third, as previously stated, the data consist of a small number of patients, for which most either do not die or do not have a CO. Increasing the number of patient records such that more of these kinds of patients appear in the data can increase the quality of the classification algorithms and the forecasting quality of these systems, leading to better treatment. 
%
%
%


\section*{ACKNOWLEDGEMENTS}
We thank the Alexander von Humboldt (AvH) Stiftung for supporting our research. 
Dr. Schütt is supported by the Deutsche Forschungsgemeinschaft (German Research Foundation; TRR 219; Project-ID 322900939 [C07]); Prof. Marx is supported by the Deutsche Forschungsgemeinschaft (German Research Foundation; TRR 219; Project-ID 322900939 [M03, M05]). Funded under the Excellence Strategy of the Federal Government and the Länder in support of the ``Computational Ecosystem for the Analysis and Clinical Application of Multi-Organ Crosstalk'' (COAT) initiative of RWTH Aachen University.

\bibliographystyle{apalike}
\bibliography{bibliography}

\begin{thebibliography}{}

\bibitem[Adriansyah et~al., 2015]{DBLP:journals/isem/AdriansyahMCDA15}
Adriansyah, A., Munoz{-}Gama, J., Carmona, J., et~al. (2015).
\newblock Measuring precision of modeled behavior.
\newblock {\em Inf. Syst. {E} Bus. Manag.}, 13(1):37--67.

\bibitem[Adriansyah et~al., 2011]{DBLP:conf/acsd/AdriansyahSD11}
Adriansyah, A., Sidorova, N., and van Dongen, B.~F. (2011).
\newblock Cost-based fitness in conformance checking.
\newblock In Caillaud, B., Carmona, J., and Hiraishi, K., editors, {\em International Conference on Application of Concurrency to System Design}, pages 57--66. {IEEE} Computer Society.

\bibitem[Almeida et~al., 2020]{almeida2020computational}
Almeida, J.~R., Freire, P., Fajarda, O., et~al. (2020).
\newblock A computational platform for heart failure cases research.
\newblock In {\em HEALTHINF}, pages 601--608.

\bibitem[Artola et~al., 2019]{artola2019impact}
Artola, G., Larburu, N., {\'A}lvarez, R., et~al. (2019).
\newblock The impact of environmental factors on heart failure decompensations.
\newblock In {\em HEALTHINF}, pages 51--58.

\bibitem[Back et~al., 2020]{DBLP:conf/biostec/BackMH20}
Back, C.~O., Manataki, A., and Harrison, E.~M. (2020).
\newblock Mining patient flow patterns in a surgical ward.
\newblock In {\em HEALTHINF}, pages 273--283. {SCITEPRESS}.

\bibitem[Benevento et~al., 2022]{DBLP:conf/icpm/Benevento0ABPBA22}
Benevento, E., Pegoraro, M., Antoniazzi, M., et~al. (2022).
\newblock Process modeling and conformance checking in healthcare: {A} {COVID-19} case study - case study.
\newblock In {\em Process Mining Workshops - {ICPM}}, pages 315--327. Springer.

\bibitem[Bottrighi et~al., 2018]{bottrighi2018general}
Bottrighi, A., Piovesan, L., Terenziani, P., et~al. (2018).
\newblock A general framework for the distributed management of exceptions and comorbidities.
\newblock In {\em HEALTHINF}, pages 66--76.

\bibitem[Buijs et~al., 2014]{DBLP:journals/ijcis/BuijsDA14}
Buijs, J. C. A.~M., van Dongen, B.~F., and van~der Aalst, W. M.~P. (2014).
\newblock Quality dimensions in process discovery: The importance of fitness, precision, generalization and simplicity.
\newblock {\em Int. J. Cooperative Inf. Syst.}, 23(1).

\bibitem[Carmona et~al., 2018]{DBLP:books/sp/CarmonaDSW18}
Carmona, J., van Dongen, B.~F., Solti, A., et~al. (2018).
\newblock {\em Conformance Checking - Relating Processes and Models}.
\newblock Springer.

\bibitem[Chin et~al., 2014]{chin2014divide}
Chin, S.~C., Zolfaghar, K., Roy, S.~B., et~al. (2014).
\newblock Divide-n-discover: Discretization based data exploration framework for healthcare analytics.
\newblock In {\em HEALTHINF}, pages 329--333. SciTePress.

\bibitem[de~Vries et~al., 2017]{DBLP:conf/biostec/VriesNGDM17}
de~Vries, G., Neira, R. A.~Q., Geleijnse, G., et~al. (2017).
\newblock Towards process mining of {EMR} data - case study for sepsis management.
\newblock In {\em HEALTHINF}, pages 585--593. SciTePress.

\bibitem[Dunn, 1964]{dunn}
Dunn, O.~J. (1964).
\newblock Multiple comparisons using rank sums.
\newblock {\em Technometrics}, 6(3):241--252.

\bibitem[Fisher, 1992]{Fisher1992}
Fisher, R.~A. (1992).
\newblock {\em Statistical Methods for Research Workers}.
\newblock Springer New York, New York, NY.

\bibitem[Guzzo et~al., 2022]{DBLP:journals/widm/GuzzoRV22}
Guzzo, A., Rullo, A., and Vocaturo, E. (2022).
\newblock Process mining applications in the healthcare domain: {A} comprehensive review.
\newblock {\em WIREs Data Mining and Knowledge Discovery}, 12(2).

\bibitem[Kerexeta et~al., 2018]{kerexeta2018predicting}
Kerexeta, J., Artetxe, A., Escolar, V., et~al. (2018).
\newblock Predicting 30-day readmission in heart failure using machine learning techniques.
\newblock In {\em HEALTHINF}, pages 308--315.

\bibitem[Kruskal and Wallis, 1952]{kruskal}
Kruskal, W.~H. and Wallis, W.~A. (1952).
\newblock Use of ranks in one-criterion variance analysis.
\newblock {\em Journal of the American Statistical Association}, 47(260):583--621.

\bibitem[Kusuma et~al., 2020]{DBLP:conf/biostec/KusumaSMJ20}
Kusuma, G.~P., Sykes, S., McInerney, C., et~al. (2020).
\newblock Process mining of disease trajectories: {A} feasibility study.
\newblock In {\em International Joint Conference on Biomedical Engineering Systems and Technologies: HEALTHINF}, pages 705--712. {SCITEPRESS}.

\bibitem[Levin et~al., 2013]{levin2013kidney}
Levin, A., Stevens, P.~E., Bilous, R.~W., et~al. (2013).
\newblock Kidney disease: Improving global outcomes (kdigo) ckd work group. kdigo 2012 clinical practice guideline for the evaluation and management of chronic kidney disease.
\newblock {\em Kidney international supplements}, 3(1):1--150.

\bibitem[Mans et~al., 2008]{DBLP:conf/mie/MansSLPCQA08}
Mans, R., Schonenberg, H., Leonardi, G., et~al. (2008).
\newblock Process mining techniques: an application to stroke care.
\newblock In {\em {MIE}}, pages 573--578. {IOS} Press.

\bibitem[Mans et~al., 2015]{DBLP:series/sbbpm/MansAV15}
Mans, R., van~der Aalst, W. M.~P., and Vanwersch, R. J.~B. (2015).
\newblock {\em Process Mining in Healthcare - Evaluating and Exploiting Operational Healthcare Processes}.
\newblock Springer Briefs in Business Process Management. Springer.

\bibitem[McDonagh et~al., 2021]{McDonagh2021}
McDonagh, T.~A., Metra, M., Adamo, M., et~al. (2021).
\newblock 2021 {ESC} guidelines for the diagnosis and treatment of acute and chronic heart failure.
\newblock {\em European Heart Journal}, 42(36):3599--3726.

\bibitem[Munoz{-}Gama et~al., 2022]{DBLP:journals/jbi/Munoz-GamaMFJSH22}
Munoz{-}Gama, J., Martin, N., Fern{\'{a}}ndez{-}Llatas, C., et~al. (2022).
\newblock Process mining for healthcare: Characteristics and challenges.
\newblock {\em J. Biomed. Informatics}, 127:103994.

\bibitem[Pegoraro et~al., 2021]{DBLP:conf/bis/0001NBAMM21}
Pegoraro, M., Narayana, M. B.~S., Benevento, E., et~al. (2021).
\newblock Analyzing medical data with process mining: {A} {COVID-19} case study.
\newblock volume 444 of {\em Lecture Notes in Business Information Processing}, pages 39--44. Springer.

\bibitem[Reisig, 1985]{DBLP:books/sp/Reisig85a}
Reisig, W. (1985).
\newblock {\em Petri Nets: An Introduction}, volume~4 of {\em {EATCS} Monographs on Theoretical Computer Science}.
\newblock Springer.

\bibitem[Romero-Gonz{\'{a}}lez et~al., 2020]{Romero_Gonz_lez_2020}
Romero-Gonz{\'{a}}lez, G., Ravassa, S., Gonz{\'{a}}lez, O., et~al. (2020).
\newblock Burden and challenges of heart failure in patients with chronic kidney disease. a call to action.
\newblock {\em Nefrolog{\'{\i}}a}, 40(3):223--236.

\bibitem[Roy and Chin, 2014]{roy2014prediction}
Roy, S.~B. and Chin, S.-C. (2014).
\newblock Prediction and management of readmission risk for congestive heart failure.
\newblock In {\em HEALTHINF}, pages 523--528. SciTePress.

\bibitem[Sattar et~al., 2021]{Sattar_2021}
Sattar, N., Lee, M. M.~Y., Kristensen, S.~L., et~al. (2021).
\newblock Cardiovascular, mortality, and kidney outcomes with {GLP}-1 receptor agonists in patients with type 2 diabetes: a systematic review and meta-analysis of randomised trials.
\newblock {\em The Lancet Diabetes \& Endocrinology}, 9(10):653--662.

\bibitem[van~der Aalst, 2016]{DBLP:books/sp/Aalst16}
van~der Aalst, W. M.~P. (2016).
\newblock {\em Process Mining - Data Science in Action, Second Edition}.
\newblock Springer.

\bibitem[Weerdt et~al., 2010]{DBLP:conf/bpm/WeerdtBVB10}
Weerdt, J.~D., Backer, M.~D., Vanthienen, J., et~al. (2010).
\newblock A critical evaluation study of model-log metrics in process discovery.
\newblock In {\em Business Process Management Workshops}, pages 158--169. Springer.

\bibitem[Zelniker et~al., 2019]{Zelniker_2019}
Zelniker, T.~A., Wiviott, S.~D., Raz, I., et~al. (2019).
\newblock {SGLT}2 inhibitors for primary and secondary prevention of cardiovascular and renal outcomes in type 2 diabetes: a systematic review and meta-analysis of cardiovascular outcome trials.
\newblock {\em The Lancet}, 393(10166):31--39.

\bibitem[Zweth et~al., 2018]{zweth2018devices}
Zweth, J., Askari, M., Spruit, M., et~al. (2018).
\newblock Devices used for non-invasive tele homecare for cardiovascular patients: A systematic literature review.
\newblock In {\em HEALTHINF}, pages 300--307. SciTePress.

\end{thebibliography}
\end{document}